\documentclass[%
 reprint,
 amsmath,amssymb,
 aps,
onecolumn,
fleqn
]{revtex4-2}

\usepackage{graphicx}% Include figure files
\usepackage{dcolumn}% Align table columns on decimal point
\usepackage{bm}% bold math
\usepackage{mathtools}
\usepackage[utf8]{inputenc}
\usepackage{newunicodechar}
\usepackage{amsmath,amssymb}
\usepackage{latexsym}
\usepackage{xcolor}
\usepackage[
  bookmarksnumbered,   % Add chapter and section numbers to bookmarks
  pdfdisplaydoctitle,  % Show the document title in the title bar
  pdfusetitle,         % Reflect \title and \author to PDF file's meta data
]{hyperref}

\usepackage[mathlines]{lineno}% Enable numbering of text and display math
\graphicspath{ {images/} }

\begin{document}
\preprint{APS/123-QED}

\title{Transportation efficiency of hydrodynamically coupled spherical oscillators in low Reynolds number fluids}% Force line breaks with \\

\author{Weiwei Su}
 \altaffiliation[The author is now belonging to ]{Department of Complexity Science and Engineering, The University of Tokyo, Kashiwa, Chiba 277-0882, Japan}%Lines break automatically or can be forced with \\
 \email{Email: su-weiwei@g.ecc.u-tokyo.ac.jp}
\affiliation{%
 Department of Mathematical Informatics, The University of Tokyo, Bunkyo City, Tokyo 113-8654, Japan
}%
 
\author{Yuki Izumida}%
 \email{Email: izumida@k.u-tokyo.ac.jp}
\author{Hiroshi Kori}%
 \email{Email: kori@k.u-tokyo.ac.jp}
\affiliation{%
 Department of Mathematical Informatics, The University of Tokyo, Bunkyo City, Tokyo 113-8654, Japan
}%
\affiliation{%
 Department of Complexity Science and Engineering, The University of Tokyo, Kashiwa, Chiba 277-0882, Japan
}%

% \collaboration{CLEO Collaboration}%\noaffiliation

\date{\today}% It is always \today, today,

\begin{abstract}
  Most bacteria are driven by the cilia or flagella, consisting of a long filament and a rotary molecular motor through a short flexible hook. The beating pattern of these filaments shows synchronization properties from hydrodynamic interactions, especially in low Reynolds number fluids. Here, we introduce a model based on simple spherical oscillators which execute oscillatory movements in one dimension by an active force, as a simplified imitation of the movements of cilia or flagella. It is demonstrated that the flow, measured by the net transportation of a test particle, is generated by a chain of oscillators and enhanced by the hydrodynamic interactions between beads, with supports from both perturbative and numerical results. Transportation efficiency also highly correlates with hydrodynamic interactions. 
  Increments of bead numbers are generally expected to produce stronger flow and efficiency, at least for small numbers of beads. 
\end{abstract}

% \keywords{Suggested keywords}%Use showkeys class option if keyword
\maketitle

%\tableofcontents

\section{Introduction}
The cilia and flagella are the fundamental organs for microorganisms as the instruments of swimming in low Reynolds number environments \cite{brennen1977fluid, ottemann1997roles} and transporting materials like protein \cite{falk2015specialized}. The undergoing mechanics in their wave-like behaviors is also a perfect paradigm for the study of active matter \cite{ramaswamy2010mechanics}. The synchronization of the beating pattern of these filaments is common \cite{friedrich2016hydrodynamic}. Usually, for example, bacteria perform a two-gait “run-and-tumble” motion in fluids \cite{berg1993random}. From repeating such motion the bacteria gain the net momentum in order to move. Thus it is worth investigating the mechanism behind such behavior in an analytical interpretation.

Some experimental studies take efforts to discover the relation between viscous hydrodynamic interaction and synchronization, as Taylor originally proposed \cite{taylor1951analysis}. Experiments on carpets of bacteria with active flagella \cite{darnton2004moving} and arrays of artificial magnetically actuated cilia \cite{coq2011collective, shields2010biomimetic, vilfan2010self} have found collective effects via hydrodynamic interactions such as complex flow patterns and collective phase shifts. Through the synchronization of filaments, it is well known that the energy dissipation of the model can be minimized \cite{michelin2010efficiency}, so one can expect that such collective effects provide sufficient influence over the performance of the oscillator model. 

The dynamic of passively induced movement generated by the flagella is a highly complex elastohydrodynamic problem \cite{lauga2015bacterial}. Therefore, in order to solve such a model, a computationally simple but still practically accurate model is required. It has been a popular idea of representing slender filaments by spheres for hydrodynamic modeling \cite{friedrich2012flagellar}. Uchida and Golestanian propose a generic criterion to reach stabilized states for two beads with arbitrary but periodic trajectory and force profile \cite{uchida2011generic}. Recently, the details of hydrodynamic effects are categorized by direct and indirect effects from the slender structure of filaments, from approximating the filament as a point mass orbiting around the cell \cite{chamolly2020direct}.

One representative of a minimal model can be found in Ref.~\cite{uchida2012hydrodynamic}. In this literature, a pair of spherical beads orbiting in the same shape of periodic trajectory near a wall is investigated. At first, the linear stability of the synchronized state is analyzed and examined by, for instance, circular, linear, and elliptical trajectories.
%for the stability analysis of the synchronized state.
Furthermore, the analysis of nonlinear evolution for these trajectories is done in the far-field limit, as well as their near-field interaction between two beads due to the finite size of the trajectory. 

For a better comprehension of the low Reynolds number hydrodynamics, there are literatures such as Kotar et al. \cite{kotar2013optimal}, which focuses on the optimal condition of hydrodynamic synchronization for rotors in circular trajectories. This article numerically examines the model setups proposed by Niedermayer et al. \cite{niedermayer2008synchronization} and Uchida \& Golestanian \cite{uchida2011generic}, that a smaller spring stiffness or larger asymmetry leads to stronger coupling condition and synchronization. Another conclusion is, if some thermal noise is included in the original setup, the synchronization still holds for not very high temperatures. Liao \& Lauga \cite{liao2021energetics} have examined a model of two spherical bodies in circular orbits of late as well, and the condition of minimal energy dissipation rate is discovered. 

In this article, we build a simple model, namely, a chain of spherical beads that oscillate in a one-dimensional orbit and are hydrodynamically coupled with one another, and focus on the transportation efficiency of the system, which is rarely focused on by previous literature. Our results are expected to contribute to describe the ciliary motion in mucociliary transport of the respiratory system and oviduct transport of the female reproductive system \cite{sleigh1988propulsion, kuek2020first, bianchi2021control, osterman2011finding}. Moreover, the metachronal pattern we discovered in the case of many beads provides theoretical implications for some motion generated from motile cilia in bacteria like Paramecium cells \cite{narematsu2015ciliary}. 

The flow of text is as the following: We will first introduce a minimal model, with only two beads inside. In the two-bead model, we provide the theoretical analysis of the dynamical system using a perturbation method, yielding expressions for net transportation, work, and efficiency.
We then numerically analyze the case of many beads, indicating the positive cooperative effect via hydrodynamic interaction between the beads.

\section{The general setup} \label{s2}
\begin{figure}[!htbp]
    \centering
    \includegraphics[width=.4\linewidth]{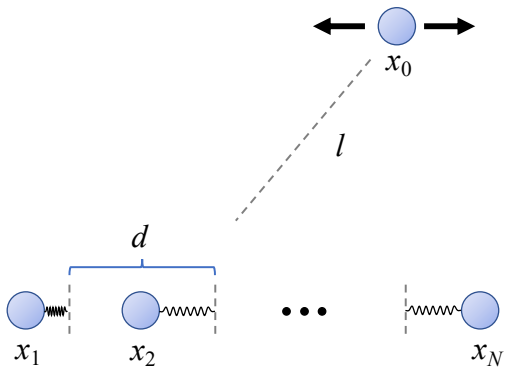}
    \caption{The setup of spherical bead model. Here, $d$ is the distance between the equilibrium positions of neighboring beads; $l$ is the typical distance between the beads and the test particle.
} \label{fig221}
\end{figure}
\begin{figure}[!htbp]
    \centering
  \includegraphics[width=0.8\linewidth]{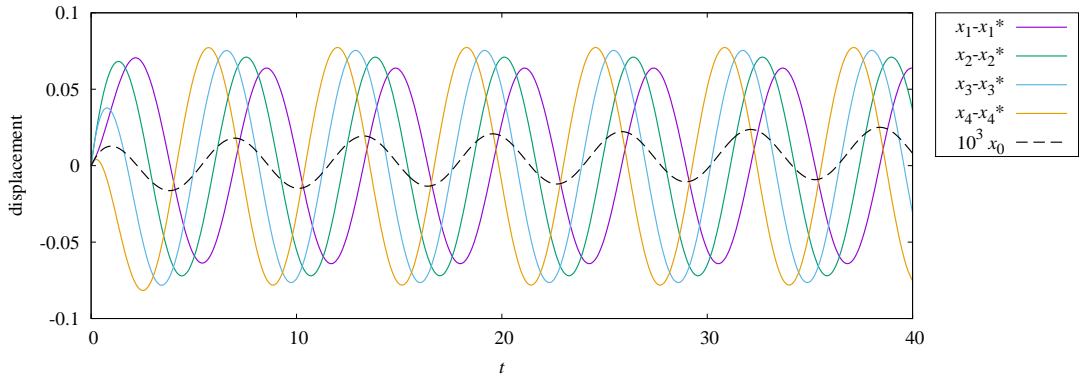}
     \caption{
  Typical dynamical behavior of the system for $N=4, \psi_i=1.0 (i-1), d=1$ and $\epsilon=0.1$.
  The displacements of the beads, $x_i-x_i^*$ ($i=1,2,3,4$), are plotted together with 
  that of the test particle, $10^3 x_0$ (instead of $x_0$ for visibility).
  The amplitudes of beads are different because of the effect of hydrodynamic coupling.
  The slope of the envelope of $x_0$ is approximately $R/T$.} 
   \label{fig:ts}
\end{figure}

As an analytically tractable model of a flow generator, a one-dimensional setup is established and illustrated in Fig. \ref{fig221}. Specifically, consider $N$ spherical beads that can move only along the $x$-axis, obeying
\begin{align}
\label{eq1}
\gamma \left[ \dot{x}_i - v (x_i) \right] = F_i,
\end{align}
where $\gamma = 6 \pi \mu a$ is the drag coefficient for the fluid viscosity $\mu$, $a$ is the radius of the beads, $x_i$ ($i= 1, 2, \ldots, N)$ is the position of bead $i$, $v(x_i)$ is the flow field at $x_i$ generated by the beads $j \neq i$, and $F_i$ is the active force applied to bead $i$. The flow field is given by
\begin{align}
    v (x_i) \coloneqq \sum_{j \neq i} H_{ij} F_j, 
\end{align}
where $H_{ij}$ $(1\leq i,j \leq N)$ is the Oseen tensor for the Stokeslet in incompressible Newtonian fluid \cite{dhont1996introduction}. Because only one-dimensional motion in the $x$-direction is considered, the Oseen tensor is given as
\begin{align} \label{eq2_2}
    H_{ij} = \frac{1}{\gamma} \frac{3a}{2|x_{ij}|} + O\left[ \left(\frac{a}{d}\right)^3 \right],
\end{align}
where $x_{ij}=x_i-x_j$ and $d$ is the distance between the equilibrium positions of the neighboring beads. Hereafter, all $O\left[ \left(\frac{a}{d}\right)^3 \right]$ terms are neglected.
Also consider a test particle of radius $a$, obeying
\begin{align}
\label{eq4}
\gamma \left[ \dot{x}_0 - v (x_0) \right] = 0, 
\end{align}
where
\begin{align}
   v(x_0) \coloneqq \sum_{j=1}^N H_{0j} F_j,
\end{align}
$v (x_0)$ is the velocity profile of the test particle.
Denote the typical distance between the flow generator and the test particle by $l$.
Here, we assume $l \gg d$ and employ the following approximation throughout this study:
\begin{align}
    H_{0j} = \frac{1}{\gamma} \frac{3a}{2l}.
\end{align}
With this approximation, $l$ only affects the magnitude of displacement of $x_0$ and can be chosen arbitrarily. 
Active force $F_i$ is modeled as follows.
Assume that each bead is connected to a pinned point by an elastic beam,
which exerts force by periodically changing its natural length.
Specifically, consider
\begin{gather}    
 F_i = k \left[ x_i^* + L(\phi_i) - x_i\right], \label{eq8}\\
 L(\phi_i)  = \epsilon L_0 \sin{\phi_i}, \label{eq9}\\
 \phi_i = \omega t + \psi_i
\end{gather}
where $x_i^* + L(\phi_i)$ is the equilibrium position, $L(\phi_i)$ is the change of the natural length, $\phi_i$ is the phase of beating, $k$ is the spring constant,
$\psi_i$ is the phase offset, and $\epsilon L_0$ is the oscillation amplitude of the natural length, $\epsilon$ is a nondimensional quantity introduced for later convenience.
Without loss of generality, set $x_i^*=(i-1)d$ and $\psi_1=0$.
One can interpret that beams repeat contraction and expansion with period $T=\frac{2\pi}{\omega}$
 with some phase differences between the neighbors. 
In the following numerical simulations, without specific mentions we fix numerical values of parameters as: $a=0.1$, $\gamma=1, l=1000, L_0=1, \omega=1$ and vary $N, d, \epsilon$ and $\psi_i$ $(i=2,\ldots,N)$. Typical behavior for $N=4$ is displayed in Fig.~\ref{fig:ts}.

To quantify the system performance, 
define the following efficiency for the transportation of the test particle:
\begin{align} \label{eq_eff_alter_1}
\hat{\eta} \coloneqq \frac{\gamma \big<v_0 \big>^2}{P}
= \frac{\gamma R^2}{W T},
\end{align}
where $\big< \cdot \big>$ denotes the average over one cycle,
$P$ and $W$ are the total power and work exerted by the flow generator for one cycle, respectively, 
and $R$ is
the net transportation of the test particle for one period.
More precisely, define
\begin{align}
 R \coloneqq \lim_{n \to \infty} \int_{n T}^{(n + 1) T} \dot x_0 \ dt 
 = \lim_{n \to \infty} \int_{n T}^{(n + 1) T} v(x_0) \ dt 
 = \lim_{n \to \infty} \frac{3a}{2 \gamma l} \int_{n T}^{(n + 1) T} \sum_{j=1}^N F_j \ dt 
\label{eq16},
\end{align}
where $n$ is the number of cycles and consider a large $n$ limit to eliminate the transient behavior. Also define
\begin{align} \label{W_define}
 W \coloneqq \sum_{i = 1}^{N} W_i,
\end{align}
where
\begin{align} \label{eq13}
W_i \coloneqq \lim_{n \to \infty} \int_{n T}^{(n + 1) T} \big(\dot{x}_i - v (x_i) \big) F_i \ dt
 = \lim_{n \to \infty} \frac{1}{\gamma}\int_{n T}^{(n + 1) T} F_i^2 \ dt.
\end{align}
One may evaluate $R$, $W$, and thus $\hat \eta$ if the time-asymptotic behavior of $x_i(t)$ $(i=1,\ldots, N)$ is obtained.

We are interested in the cooperative effect on the efficiency and thus later compare the efficiency for different $N$ values. Thus, it is convenient to introduce the per-unit efficiency:
\begin{align}
 \eta = \frac{\hat \eta}{N} = \frac{\gamma R^2}{W T N}.
\end{align}
A positive cooperative effect is indicated if $\eta$ significantly increases with $N$. 

A similar kind of efficiency could be found, such as the Lighthill hydrodynamic efficiency (cf. Ref.~\cite{lighthill1952squirming}) which happens to be one of the most popular among literatures (for instances, in Ref.~\cite{leshansky2007frictionless, golestanian2008analytic}). There, the particle is approximately spherical and deformable in order to move by morphing and the expression is 
\begin{align} \label{eq_eff_alter_2}
\hat{\eta}' \coloneqq \frac{\gamma \big<v_0 \big>^2}{\big<P \big>}
\end{align}
where $v_0$ is the velocity of the particle and $P$ is the work done from morphing.
This is fundamentally the same definition as ours if one replaces the work done from morphing to the work done by beads.

\section{The minimal model: a 2-bead case} \label{sec22}
\begin{figure}[!htbp]
    \centering
    \includegraphics[width=0.3\linewidth]{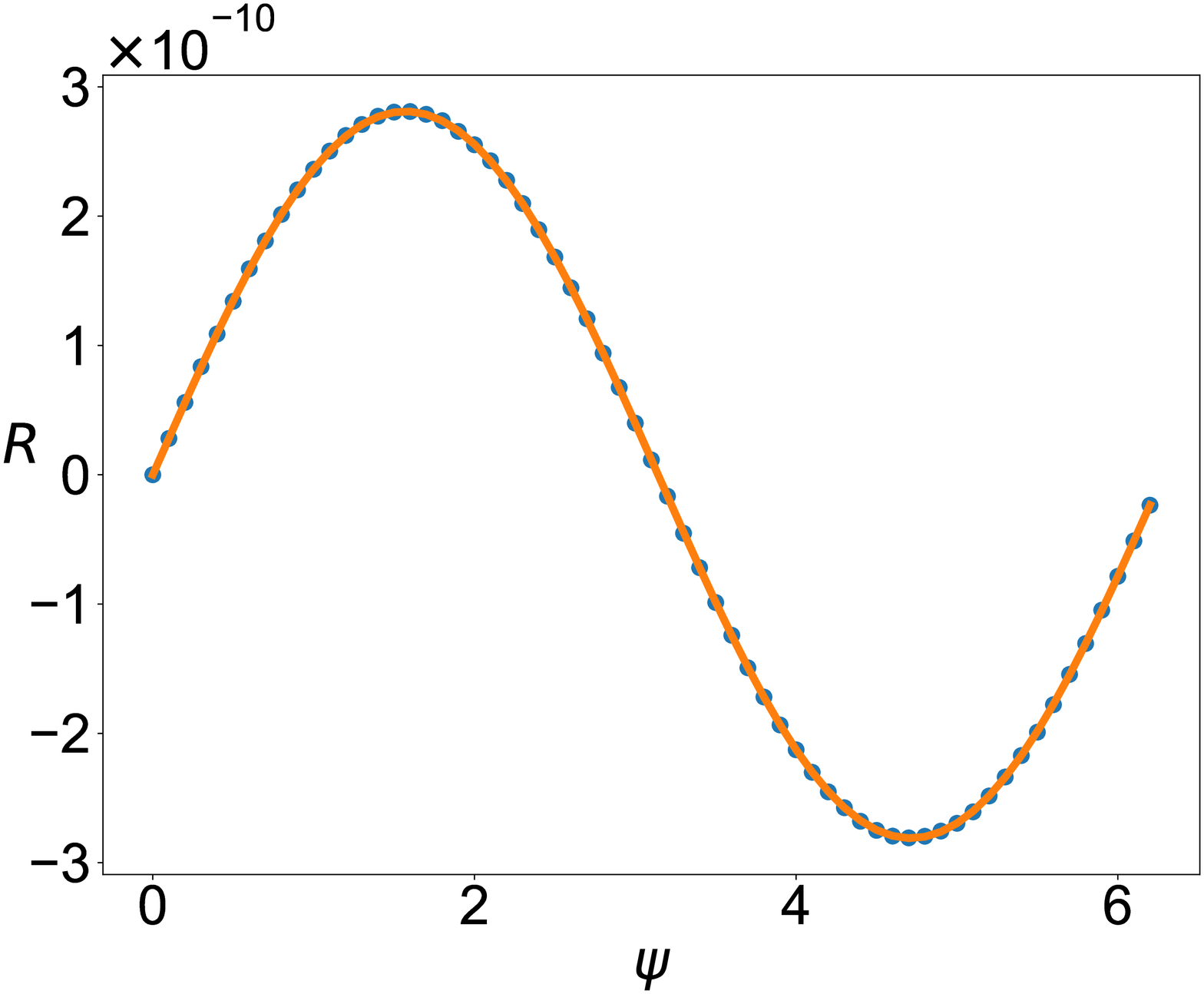}
    \includegraphics[width=0.33\linewidth]{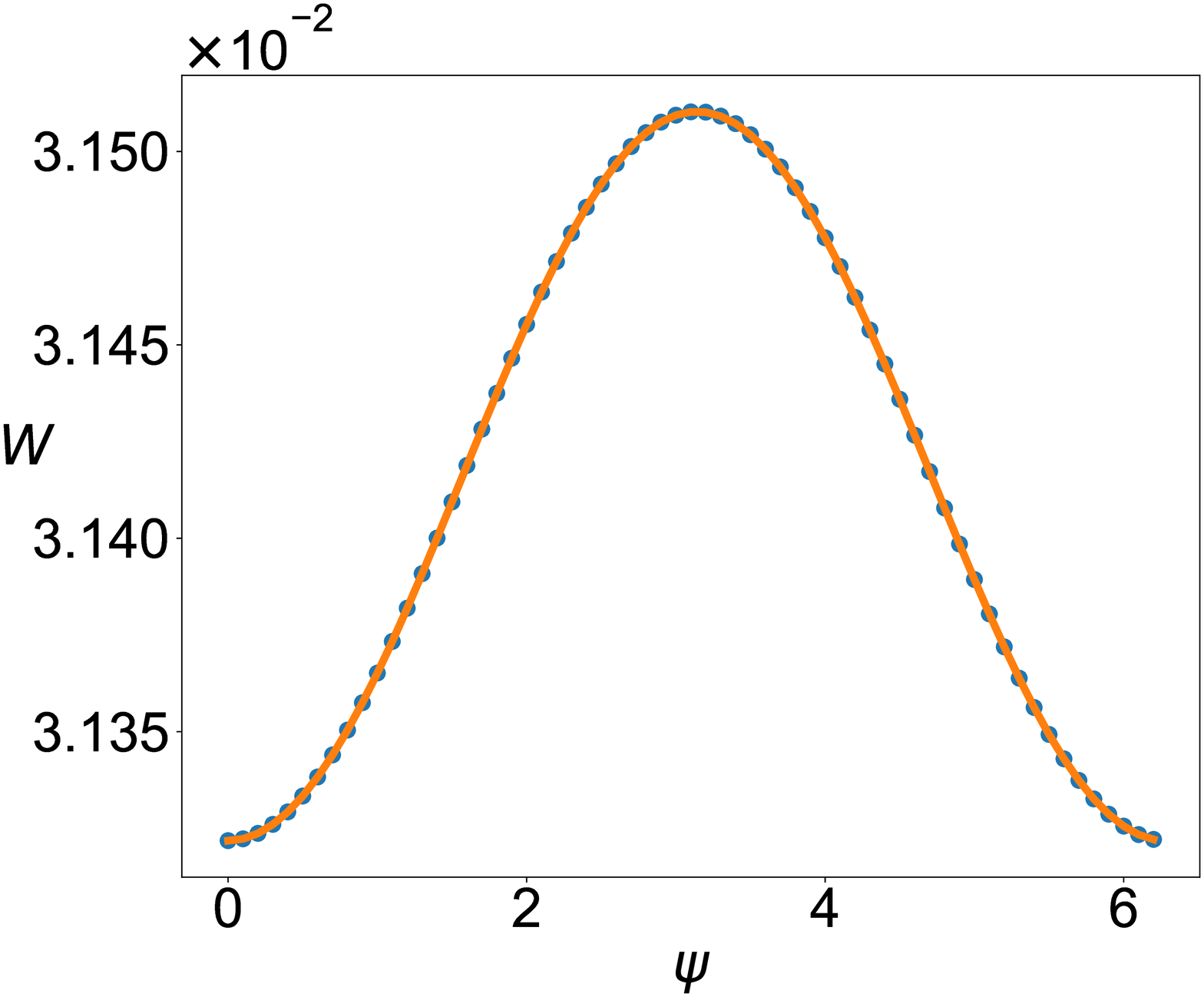}
    \includegraphics[width=0.3\linewidth]{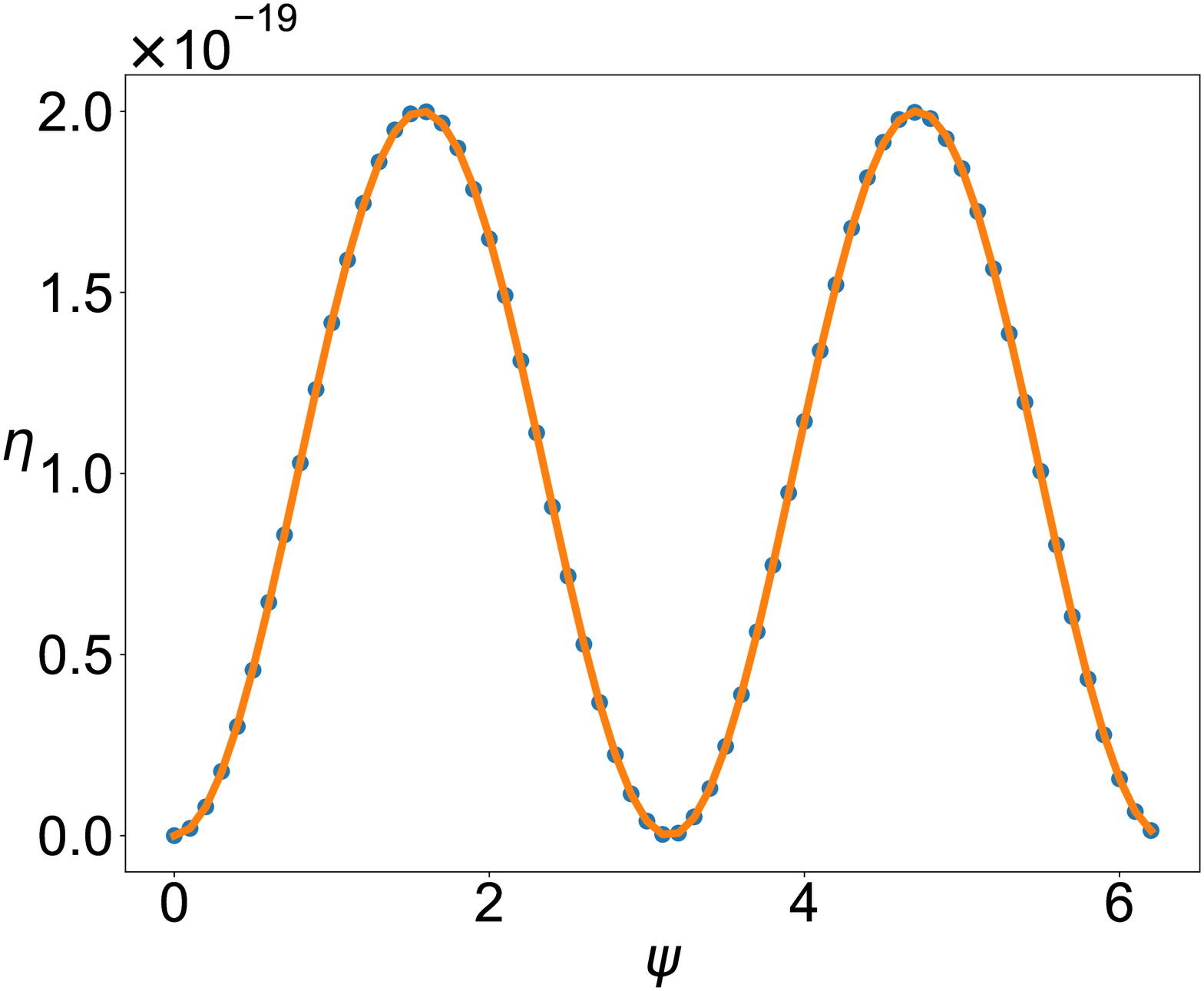}
    \caption{
The net transportation $R$ and the work $W$ over one cycle as well as the efficiency $\eta$ with respect to the phase difference $\psi$ in the two-bead case $(N=2)$.
Solid curves are the theoretical ones given in Eq.~\eqref{R_last}--\eqref{eta_last} while blue dots are the simulation results of Eq.~\eqref{eq1} and Eq. \eqref{eq4}.
Parameter values: $d=50$, $l=1000$, and $\epsilon=0.1$. 
}
    \label{fig231}
\end{figure}

Here, by considering the minimal model, the flow generator consisting of only two beads $(N=2)$,
and perturbatively solving such system, we derive the expressions for $R, W$, and $\hat \eta$. 
For simplicity, $\psi_2$ is denoted as $\psi$ in this section; thus, $\phi_1=\omega t$ and $\phi_2 = \omega t + \psi$.
Then, our model for $N=2$ is
\begin{subequations} % 2023-01-28 17:03の式群
 \label{two_bead_model}
\begin{align}
    \dot{x}_1 = \frac{3a}{2 \gamma |x_2-x_1|} F_2 + \frac{1}{\gamma} F_1, \label{eq23}\\
    \dot{x}_2 = \frac{3a}{2 \gamma |x_2-x_1|} F_1 + \frac{1}{\gamma} F_2. \label{eq24}
\end{align}
\end{subequations}
The perturbation method might be applied by treating $\epsilon$ as a small parameter. Firstly introduce the dimensionless version of canonical basis as
\begin{align}
    X := \frac{x_1 + x_2}{d}, \label{eq21}\\
    Y := \frac{x_2 - x_1}{d}. \label{eq22}
\end{align}
For simplicity, always assume $Y > 0$,
which is a reasonable assumption physically, meaning that the beads never overlap with each other, so no collision occurs.
Then, Eq.~\eqref{two_bead_model} becomes
\begin{subequations} % 2023-01-28 17:04の式群
 \label{two_bead_model2}
\begin{align}
    \dot{X} & = \frac{k}{\gamma} \left(\frac{3a}{2dY} + 1\right)
\left[1 + \epsilon \hat L_0 (\sin{\phi_1} + \sin{\phi_2}) - X \right], \label{eq25_1}\\
    \dot{Y} & = \frac{k}{\gamma} \left(-\frac{3a}{2dY} + 1\right)
\left[ 1 + \epsilon \hat L_0 (\sin{\phi_2} -  \sin{\phi_1}) - Y \right], \label{eq26_1}
\end{align}
\end{subequations}
where $\hat L_0 := \frac{L_0}{d}$.
Now, solve this using the perturbative forms:
\begin{align} 
    X(t) & = 1 + \epsilon X_1(t) + \epsilon^2 X_2(t) + O(\epsilon^3), \label{eq35}\\
    Y(t) & = 1 + \epsilon Y_1(t) + \epsilon^2 Y_2(t) + O(\epsilon^3). \label{eq36}
\end{align}
Note
\begin{align} \label{eq37}
    \frac{1}{Y}
    = 1 - \epsilon Y_1 + \epsilon^2 (- Y_2 + Y_1^2) + O(\epsilon^3).
\end{align}
Therefore, by collecting $O(\epsilon)$ in Eq.~\eqref{two_bead_model2}, 
\begin{subequations} % 2023-01-28 17:08の式群
 \label{O1}
\begin{align}
\dot{X}_1 = {} & \alpha \left[ m(t) - X_1 \right], \\
\dot{Y}_1 = {} & \beta \left[ h(t) - Y_1 \right], 
\end{align}
\end{subequations}
where $\alpha := \frac{k}{\gamma} (\frac{3a}{2 d} + 1)$, $\beta := \frac{k}{\gamma} (- \frac{3a}{2 d} + 1)$, $m(t) := \sin{\phi_1} + \sin{\phi_2}$, and $h(t) := \sin{\phi_2} - \sin{\phi_1}$. 
Similarly, for $O(\epsilon^2)$, 
\begin{subequations} % 2023-01-28 17:08の式群
 \label{O2}
\begin{align}
    \dot{X}_2 & = - \frac{3 a k Y_1}{2 \gamma d} \left[ m(t) - X_1 \right] - \alpha X_2, \\
    \dot{Y}_2 & = \frac{3 a k Y_1}{2 \gamma d} \left[ h(t) - Y_1 \right] - \beta Y_2. 
\end{align}
\end{subequations}
Solving these equations, the time-asymptotic expressions are obtained as 
\begin{subequations} \label{eq:xandy} % 2023-01-28 17:17の式群 
\label{expressions}
\begin{align}
 {X}_1 = & \frac{2 \alpha  \cos \left(\frac{\psi }{2}\right)
\left[\alpha  \sin \left(\omega t+\frac{\psi }{2}\right)-
 \omega  \cos \left(\omega t +\frac{\psi }{2}\right)\right]}{\alpha ^2+\omega ^2}, \\
 {Y}_1 = & \frac{2 \beta  \sin \left(\frac{\psi }{2}\right)
\left[ \beta  \cos \left(\omega t +\frac{\psi }{2}\right)
 +\omega  \sin \left(\omega t +\frac{\psi }{2}\right)\right]}{\beta ^2+\omega ^2},\\
 {X}_2 = & -\frac{3 a \beta  k \omega  \sin (\psi ) }
 {2 \alpha \gamma  d \left(\alpha ^2+\omega ^2\right) \left(\alpha ^2+4 \omega ^2\right)
\left(\beta ^2+\omega ^2\right)} \nonumber \\
 & \left[\left(\alpha ^2+4 \omega ^2\right) \left(\alpha  \beta
 +\omega ^2\right)+\alpha  \omega  \left(\alpha ^2+3 \alpha  \beta -2 \omega ^2\right)
 \sin (2 \omega t +\psi )
    +\alpha  \left(\alpha ^2 \beta -3 \alpha  \omega ^2-2 \beta  \omega ^2\right)
 \cos (2 \omega t +\psi ) \right],\\
    {Y}_2 = & -\frac{3 a \beta  k \omega  \sin ^2\left(\frac{\psi }{2}\right) \left[\left(2 \omega ^3-4 \beta ^2 \omega \right) \cos (2 \omega t +\psi )+\beta  \left(\beta ^2-5 \omega ^2\right) \sin (2 \omega t +\psi )\right]}{\gamma d \left(\beta ^2+\omega ^2\right)^2 \left(\beta ^2+4 \omega ^2\right)}.
\end{align}
\end{subequations} 
%\begin{align}
%    {X}_1 = & \frac{2 \alpha  \cos \left(\frac{\psi }{2}\right) \left[\alpha  \sin \left(\omega t+\frac{\psi }{2}\right)-\omega  \cos \left(\omega t +\frac{\psi }{2}\right)\right]}{\alpha ^2+\omega ^2}, \\
%    {Y}_1 = & \frac{2 \beta  \sin \left(\frac{\psi }{2}\right) \left(\beta  \cos \left[\omega t +\frac{\psi }{2}\right)+\omega  \sin \left(\omega t +\frac{\psi }{2}\right)\right]}{\beta ^2+\omega ^2},
%\end{align}
%%while ${X}_2$ and ${Y}_2$ are
%\begin{align}
%    & \begin{aligned}
%    {X}_2 = & -\frac{3 a \beta  k \omega  \sin (\psi ) }{2 \alpha \gamma  d \left(\alpha ^2+\omega ^2\right) \left(\alpha ^2+4 \omega ^2\right) \left(\beta ^2+\omega ^2\right)} \\
%    & \Big(\left(\alpha ^2+4 \omega ^2\right) \left(\alpha  \beta +\omega ^2\right)+\alpha  \omega  \left(\alpha ^2+3 \alpha  \beta -2 \omega ^2\right) \sin (2 \omega t +\psi ) \\
%    & \ \ \ \ +\alpha  \left(\alpha ^2 \beta -3 \alpha  \omega ^2-2 \beta  \omega ^2\right) \cos (2 \omega t +\psi ) \Big),
%    \end{aligned}
%    \\
%    & \ {Y}_2 = -\frac{3 a \beta  k \omega  \sin ^2\left(\frac{\psi }{2}\right) \left[\left(2 \omega ^3-4 \beta ^2 \omega \right) \cos (2 \omega t +\psi )+\beta  \left(\beta ^2-5 \omega ^2\right) \sin (2 \omega t +\psi )\right]}{\gamma d \left(\beta ^2+\omega ^2\right)^2 \left(\beta ^2+4 \omega ^2\right)}.
%\end{align}
With these expressions, one may now calculate $R$ and $W$.
Substituting Eq.~\eqref{expressions} into Eq.~\eqref{eq16}, 
\begin{align}
\label{R_last}
 R %= \frac{3a}{2\gamma l} \int_0^T (F_1 + F_2) dt
% &= \frac{3a k d}{2\gamma l} \int_0^T
%\left(1 + \epsilon \sin (\omega t +\psi )+\epsilon \sin (\omega t)-\left(1 + \epsilon X_1 + \epsilon ^2 X_2 + O(\epsilon^3) \right)\right) \ dt \nonumber \\
 = \frac{9 \pi  a^2  k^2 \sin (\psi ) \beta \left(\alpha  \beta +\omega ^2\right)}{2  l \gamma^2 \alpha \left(\alpha ^2+\omega ^2\right) \left(\beta ^2+\omega ^2\right)} \epsilon ^2 \left[1+O(\epsilon)\right]
\end{align}
which vanishes in the first order of $\epsilon$ while generally not in the second order of $\epsilon$ (cf. Eq.~\eqref{eq_eff_alter_1} and Eq.~\eqref{eq:xandy}). Substituting Eq.~\eqref{expressions} into Eq.~\eqref{W_define}
and using $F_1^2 + F_2^2 = \frac{(F_1+F_2)^2 + (F_1-F_2)^2}{2}$, 
\begin{align}
 \label{W_last}
    W
%= & \int_{0}^T \frac{k^2 d^2}{2 \gamma} \bigg[\Big(\epsilon \big(\sin{\phi_1} + \sin{\phi_2} - X_1(t) \big) - \epsilon^2 X_2(t) - O(\epsilon^3) \Big)^2 
%    + \Big(\epsilon \big(\sin{\phi_2} - \sin{\phi_1} - Y_1(t) \big) - \epsilon^2 Y_2(t) - O(\epsilon^3) \Big)^2 \bigg] \ dt \\
    = & \frac{\pi  k^2 \omega d^2 \left[ \left(\beta ^2-\alpha ^2\right) \cos (\psi )+\alpha ^2+\beta ^2+2 \omega ^2\right]}{\gamma  \left(\alpha ^2+\omega ^2\right) \left(\beta ^2+\omega ^2\right)}  \epsilon ^2 \left[1+O(\epsilon)\right]
\end{align}
which implies similar minimal energy dissipation versus phase difference as in Ref. \cite{liao2021energetics}. Therefore, 
\begin{align} \label{eta_last}
    \eta &= \frac{81 a^4 \beta ^2 k^2 \sin ^2(\psi ) \left(\alpha  \beta +\omega ^2\right)^2}{16 \gamma^2 \alpha ^2 d^2 l^2 \left(\alpha ^2+\omega ^2\right) \left(\beta ^2+\omega ^2\right) \left[\left(\beta ^2-\alpha ^2\right) \cos (\psi )+\alpha ^2+\beta ^2+2 \omega ^2\right]} \epsilon ^2 
\left[1+O(\epsilon)\right].
\end{align}
It should be noted that all quantities start from $O(\epsilon^2)$ because the contributions of first-order terms are averaged out over one cycle. 
As shown in Fig. \ref{fig231}, these expressions are in excellent agreement with the simulation results. 

Furthermore, by recalling that $\frac{a}{d}$ is assumed to be small and using $\alpha=\beta=\frac{k}{\gamma}+O(\frac{a}{d})$, the expressions can be much simplified as 
\begin{subequations} % 2023-02-7 16:59の式群
\label{n2results} 
\begin{align}
R &=\frac{9 \pi  a^2  L_0^2 k^2 \sin (\psi )}{2 l d^2 (k^2+\gamma^2 \omega ^2)}
\epsilon ^2 \left[ 1+O\left(\epsilon, \frac{a}{d}\right)\right], \label{R_}\\
W &= \frac{2 \pi  L_0^2 k^2 \omega \gamma }{\left(k^2+\gamma^2 \omega ^2 \right)}
\epsilon ^2 \left[ 1+O\left(\epsilon, \frac{a}{d}\right)\right], \label{W_}\\
 \eta &= \frac{81 a^4 L_0^2 k^2 \sin ^2(\psi )}{32 l^2 d^4 \left(k^2 + \gamma^2 \omega ^2\right)}
\epsilon ^2 \left[ 1+O\left(\epsilon,\frac{a}{d}\right) \right]. \label{eta_}
\end{align}
\end{subequations}

From these expressions, observe that
$R$ decreases with increasing $d$, proportionally to $1/d^2$. Thus, $R$ vanishes at a large $d$ limit, indicating the essential role of hydrodynamic interaction on the generation of the net flow.
In contrast, $W$ is almost independent of $d$ and the leading term is just the sum of the works of two independent beads; the hydrodynamic interaction gives little contribution to the work for $d \gg a$.  Therefore, $\eta$ is approximately proportional to $R^2$ and its maxima approximately coincide with the maxima of $R$. 

%Additionally, under the limit where $a/d \to 0$, so $\beta \approx \alpha$ and the expressions could be much simplified as 

\section{Multiple bead case} \label{ch3}
\begin{figure}[tb]
    \centering
    \includegraphics[width=0.5\linewidth]{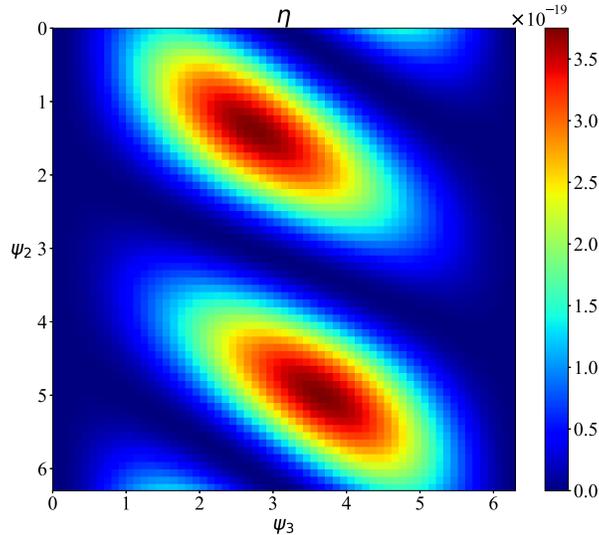}
    \caption{Efficiency $\eta$ of the three-bead system as a function of phase offsets $\psi_2$ and $\psi_3$ for $N=3$. Parameter values are the same as those in Fig.~\ref{fig231}. 
%The spectrum plot of efficiency with respect to two phases. The value of efficiency $\eta$ increases from the dark blue color to red. Every grid has an edge length of 0.1, which means the spectrum is scanned with a segment length of 0.1 for each $\psi$.
}
    \label{fig321}
\end{figure}

In this section, a more general setup is considered, where the number of bead $N$ is no longer restricted by two. In the simulation, we set $\psi_i = i \Delta \psi$, where $\Delta \psi$ is a constant describing the phase difference. In this setup, one only needs to consider $0\leq \Delta \psi \leq \pi$ because of the symmetry of the system. Such a pattern mimics the metachronal wave ubiquitously found in the collective motion of flagella or cilia (cf. Ref.~\cite{elgeti2015physics}).
For $N=3$, as shown in Fig. \ref{fig321}, this setup is numerically proven to approximately produce the maximum efficiency output for the cases of $N=3$. The figure represents the efficiency as a function of $\psi_2$ and $\psi_3$, in which the optimality is found at $(\psi_2, \psi_3)\approx (1.3, 2.6)$, thus $\psi_3 \approx 2 \psi_2$. 

\begin{figure}
 \centering
   \includegraphics[width=0.4\linewidth]{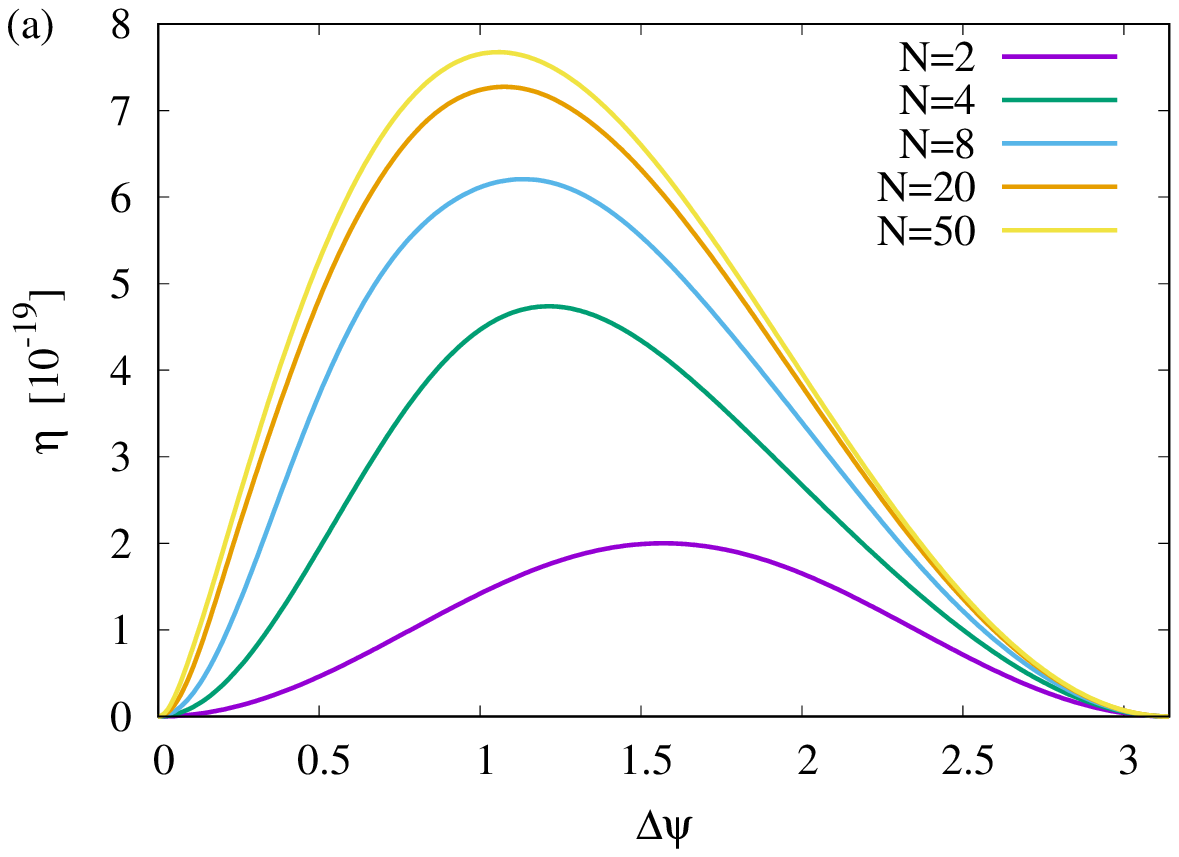}
 \hspace{5mm}
   \includegraphics[width=0.4\linewidth]{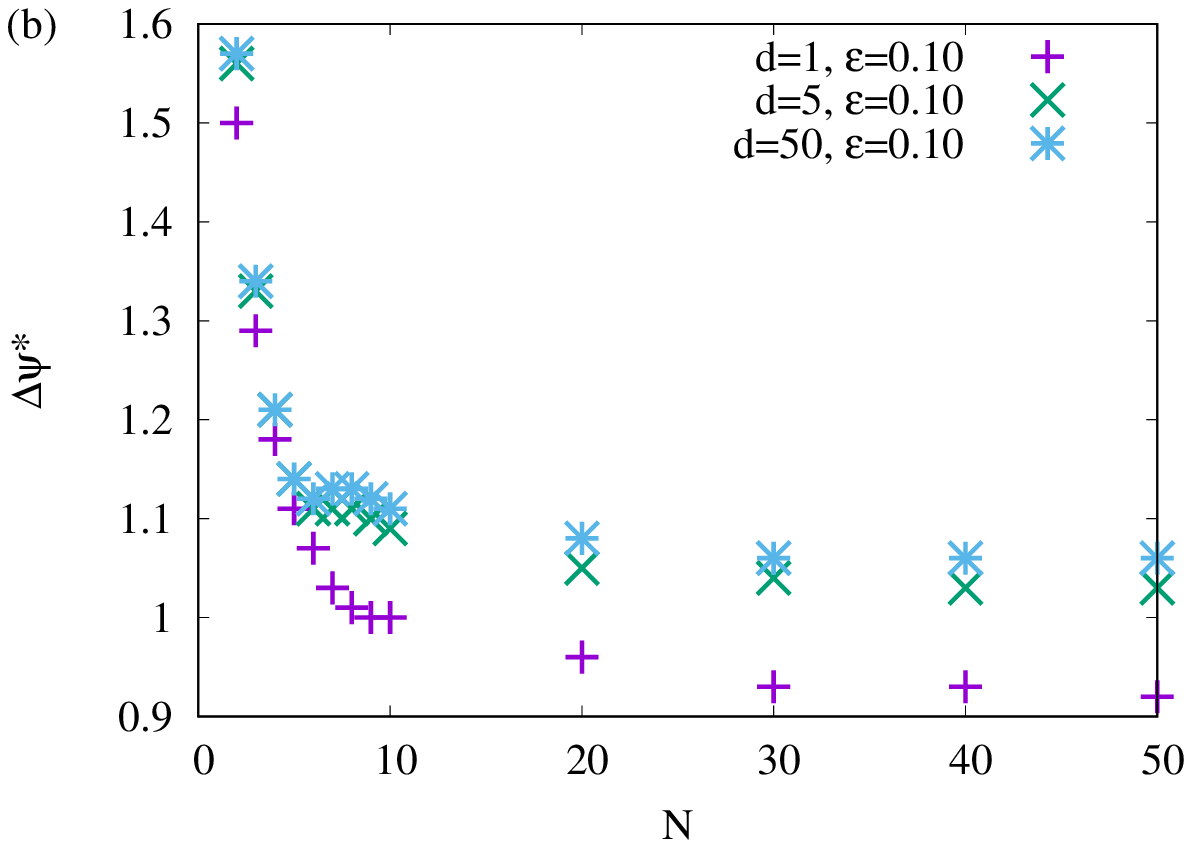}
\caption{(a) Per-unit efficiency $\eta$ versus phase difference $\Delta \psi$ in the $N$-beads cases. (b) Optimal phase difference} $\Delta \psi^*$ versus $N$. 
\label{fig:eta} 
\end{figure}

We first observe the dependence of per-unit efficiency $\eta$ on $\Delta \psi$, shown in Fig.~\ref{fig:eta}(a). It is seen that the optimal phase difference, denoted as $\Delta \psi^*$, decreases as $N$ increases. In Fig.~\ref{fig:eta}(b), we further plot the dependence of $\Delta \psi^*$ on the bead number $N$, suggesting that $\Delta \psi^*$ approaches a certain value as $N$ increases. There, we test two different $d$ values and find that the dependence on $d$ is rather weak.

\begin{figure}
 \centering
   \includegraphics[width=0.3\linewidth]{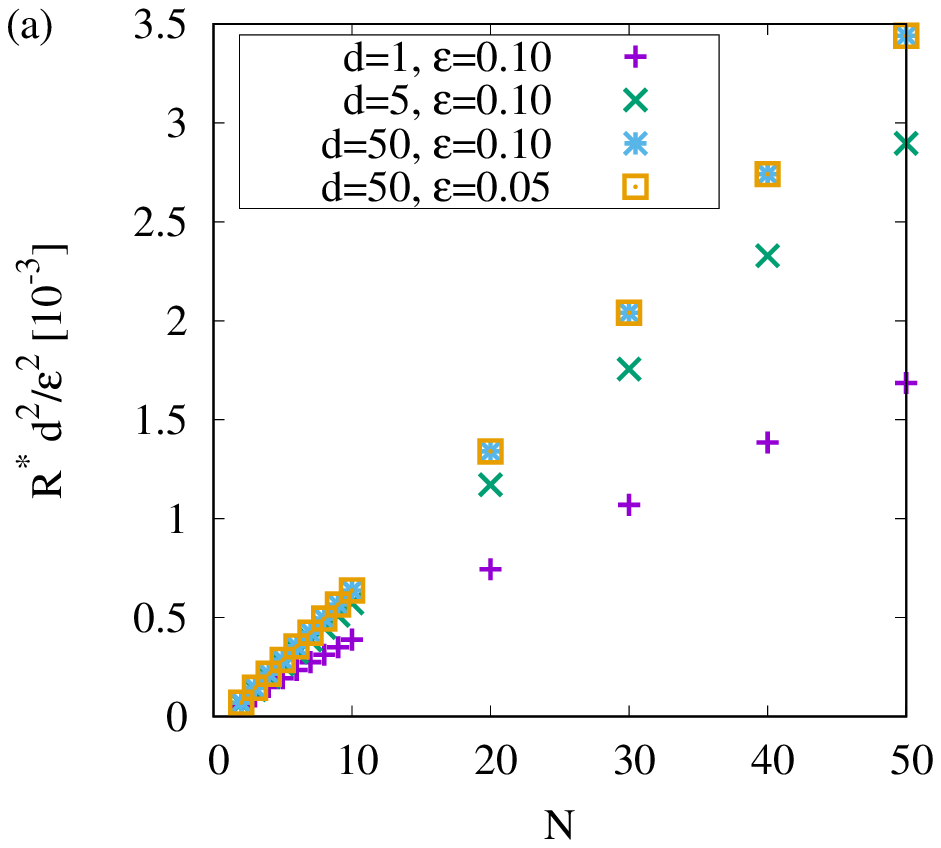}
   \includegraphics[width=0.3\linewidth]{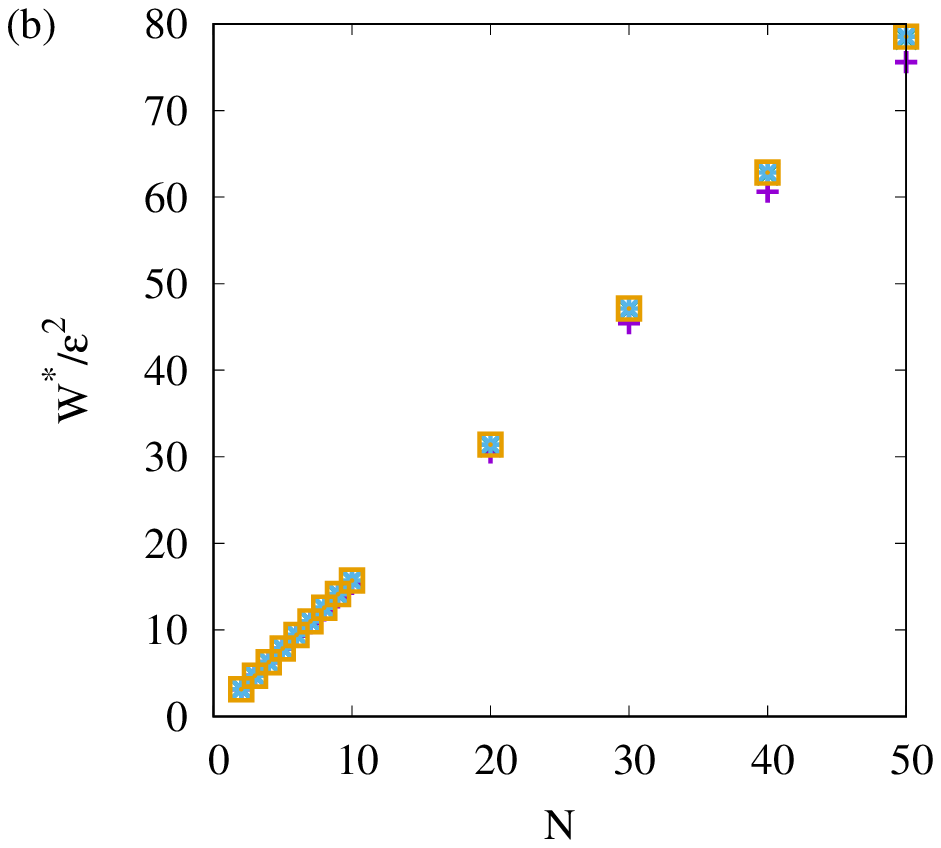}
   \includegraphics[width=0.3\linewidth]{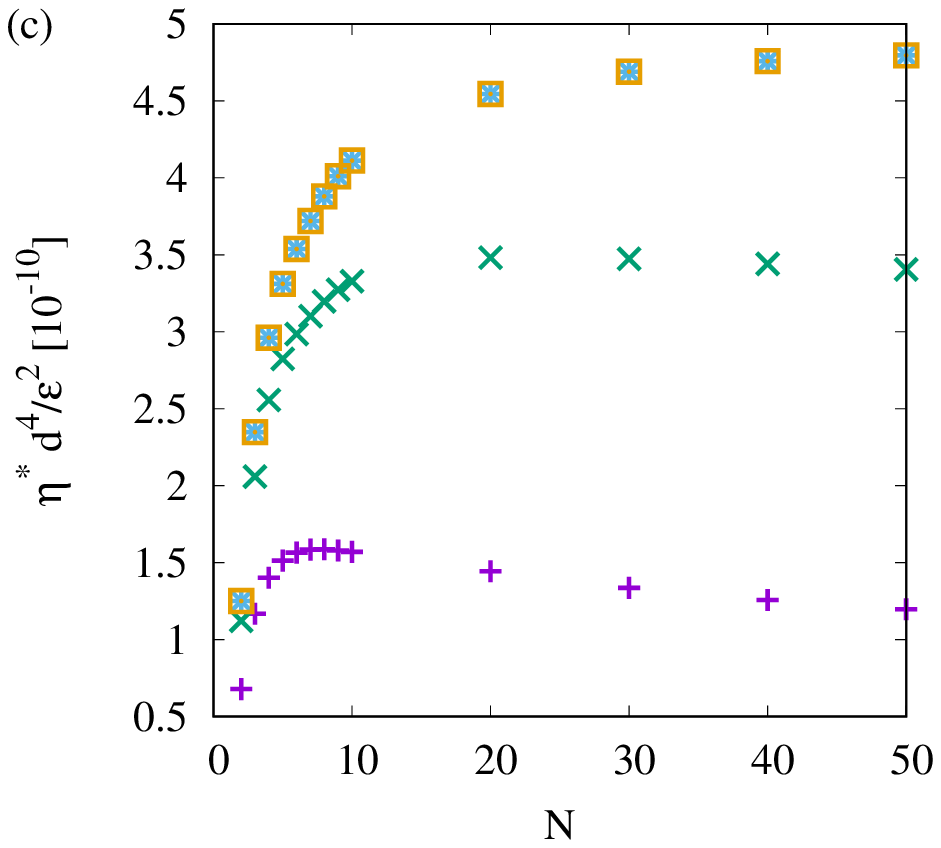}
\caption{Net transportation $R^*$, work $W^*$, and per-unit efficiency $\eta^*$ for optimal phase difference $\Delta \psi^*$ as functions of $N$.}
\label{fig:nbody} 
\end{figure}

We then observe the $N$-dependencies of $R, W$ and $\eta$ values for $\Delta \psi=\Delta \psi^*$, denoted as $R^*, W^*$, and $\eta^*$, respectively. We plot these values in Fig.~\ref{fig:nbody}, where the vertical axes are scaled as $R^* \frac{d^2}{\epsilon^2}, W^* \frac{1}{\epsilon^2}$, and $\eta^* \frac{d^4}{\epsilon^2}$. This scaling is motivated from the theoretical predicted dependence on $d$ and $\epsilon$ for $N=2$, given in Eq.~\eqref{n2results}; i.e., $R \propto \frac{\epsilon^2}{d^2}, W \propto \epsilon^2$, and $\eta \propto \frac{\epsilon^2}{d^4}$. 
Figure~\ref{fig:nbody}(a) indicates that $R$ increases approximately linearly with $N$, particularly for large $d$. Moreover, as shown in Fig.~\ref{fig:nbody}(b), $W^*$ increases virtually linearly with $N$, indicating that the work was approximately equal to the summation of independent beads. 
Assuming $R, W \propto N$, $\eta=\frac{R^2}{WN}$ should be independent of $N$; however, as shown in Fig.~\ref{fig:nbody}(c), $\eta$ is actually strongly dependent on $N$, indicating that the growth of $R$ significantly deviates from a linear growth.
Importantly, $\eta$ for $N>2$ is larger than $\eta$ for $N=2$, indicating the positive cooperative effect of hydrodynamic coupling.
It should also be noted that $\eta$ saturates or even turns to decrease for large $N$, which could come from our simple setting for the phase offsets and can possibly be improved by considering a more complex pattern of $\psi_i$. 
It is also observed that all the quantities at given $N$ in Fig.~\ref{fig:nbody} have similar magnitudes, suggesting that the dependence $R \propto \frac{\epsilon^2}{d^2}, W \propto \epsilon^2$, and $\eta \propto \frac{\epsilon^2}{d^4}$ roughly holds true for $N \ge 3$.

%Numerical results are summarized in Fig.~\ref{fig:eta}. 
%In Fig.~\ref{fig:eta}(b),  In Fig.~\ref{fig:eta}(c), the maximum $\eta$ increases as $N$ increases, indicating the positive cooperative effect of hydrodynamic coupling. The data also suggests that the cooperative effect is saturating as $N$ increases.
%To check the $\epsilon$-dependency, we also present $\eta$ for smaller $\epsilon$ value in Fig.~\ref{fig:eta}(c), indicating $\eta \propto \epsilon^2$ for $N \ge 3$ as well as $N=2$. 

%For $N=2$, we set $\Delta \psi(2) = \frac{\pi}{2}$, which is the optimal phase difference found in Sect. 3. For $N \ge 3$, we find the optimal phase difference $\Delta \psi^*(N)$ by numerically evaluating $\eta$ values for different $\Delta \psi$ values.

\section{Conclusion} \label{ch6}
We introduced a simple one-dimensional coupled sphere motor model, with the generated flow and transportation efficiency investigated both theoretically and numerically. For the minimal case with two beads, the proper theoretical expression for both the net transportation and the efficiency are derived by the perturbation method, and it is proven that these results are in excellent agreement with the numerical results. Furthermore, we numerically investigated the case of $N\ge 3$ and found that the efficiency increases with $N$.
%the increasing coupling strength give rises to the efficiency of individual bead.
%, yet such benefit is saturated for bead number close to 6.
%Moreover, to a more practical usage, the case with wall effect is examined, and the similar trend of relation with bead number for the displacement and efficiency is again found with even larger dependence. Finally, the possible autonomous design is discussed, and for the minimal case with small coupling strength, there are always two convergence phases; however, if the coupling strength increases, the asymptotic convergent pattern will gradually change to oscillatory pattern starting from the phase parameter close to the convergence phases. 

% This model can be applied to the case with, for example, an oblate trajectory of beads.
% %that the movement of beads can be approximated as a one-dimensional movement.
For a more practical usage, the effect of a wall, to which the system is pinned, 
should also be examined; such a setup can be treated by using the Blake tensor \cite{blake1972model,uchida2012hydrodynamic} instead of the Oseen tensor. It could be very helpful to derive theoretical expressions for the net transportation and efficiency for this case.
In our model, oscillators were designed non-autonomous for simplicity; 
to understand the role of synchronization and self-organized metachronal wave pattern, 
it is essential to extend the model to the case of autonomous oscillators and 
compose a theoretical framework. Such an effort would substantially contribute to
deepening the understanding of the hydrodynamic effect of interactive objects in low Reynolds numbers.
Moreover, it is worth discovering whether the higher order interactions will eventually destroy the positive coupling effect completely for very large number of oscillators, and if this is also true with effect of a wall. 
In addition, a two-dimensional movement model could be considered, as many other publications have focused, while the complexity in finding a theoretical expression could be fairly challenging.

\bibliography{ref}% Produces the bibliography via BibTeX.

\end{document}